# RAMAN SPECTROMETRY, A UNIQUE TOOL TO ANALYZE AND CLASSIFY ANCIENT CERAMICS AND GLASSES

**Ph. Colomban*[1]**

**Groupe des Nanophases et Solides Hétérogènes**
**Laboratoire Dynamique, Interactions et Réactivité**
**UMR 7075 CNRS and Université Pierre et Marie Curie**
**2 rue Henry-Dunant, 94320 Thiais, France**

**Abstract**

Raman micro/macro-spectroscopy allows for a non-destructive remote analysis: body and glaze, crystalline and amorphous phases can be identified, including the nanosized pigments colouring the glaze. Last generation instruments are portable which allows for examination in museum, on sites, etc. This paper gives an overview on the potential of Raman spectrometry technique to analyze ancient ceramics and glasses. Selected glasses as well as glazes of various porcelains, celadon, faiences and potteries, representative of the different production technologies used in the Antique, European, Mediterranean, Islamic and Asian worlds were studied. Their identification is based on the study of the Raman fingerprint of crystalline and glassy phases. Raman parameters allow for the classification as a function of composition and/or processing temperature. Special attention is given to the spectra of amorphous and coloring phases (pigment).

[1] *Tel.+33 14978 1105; Fax. +33 14978 1118; E-mail: colomban@glvt-cnrs.fr

# 1. Introduction

Leading experts generally base their certification of ancient artifacts on stylistic analysis and a personal feeling involving the five senses but more objective proofs are mandatory for identification purpose. We demonstrate the potential of Raman spectroscopy as a non-destructive technique for the characterization of ceramics and glasses [1,2]. We first present the techniques and concepts used to extract salient features from bodies, glazes and pigments of the different productions. Examples are chosen from artifacts from the beginning of the ceramic industry (Sa Huynh culture, < 3000 BP) [3], including Punic/Roman time ($2^{nd}$ BC to AD) [4], stoneware, celadon and porcelains from Vietnam (the $6^{th}$ -$16^{th}$ centuries)[3,5-8], soft- and hard-paste porcelains from Europe (18-$20^{th}$ centuries) [9-12] and faiences or pottery from the Islamic world (Ifriqiya [13], Persia, [14], Samarkand and Silk Road [15], Turkey [16] $11^{th}$-$17^{th}$ centuries). The aim of this work is to demonstrate the potential of the Raman spectroscopy for non-destructive identification of art pieces and of their technology.

## 1.1 Ceramics, glass processing and glazes

A ceramic is an artificial rock obtained by firing mixed raw materials that are more or less transformed by the thermal treatment. Ceramics are composite at the tenth of micron scale and their microstructure consists of sintered grains. Crystalline and glassy phases are present together: un-reacted, incompletely dissolved raw materials as well as some phases formed during the process are crystalline. Although a (small) part of the material is molten during the firing process of the ceramic body and form glassy phases on cooling, raw materials are almost fully molten to produce a glass or a glaze, but small crystals (< 0.1 μm to keep optical clearness) precipitate on cooling in many glasses.

Different materials are obtained if different technologies are applied to the same batch or if a given technology is applied to different batches. On the other hand, different technologies give products of very similar outward appearances (from the visual and sensory points of view), although these products are completely different in their micro/nanostructure. Much information on the processing remains written in the sample. Raman analysis of the micro-structure (for ceramics) and nano-structure (for glasses and enamels) offers a way to identify the composition, the processing and, sometimes, the age of ancient artefacts.

## 1.2. Experimental procedure – The Raman micro-/macro-spectrometry

Raman spectroscopy is an optical technique and, thus, can be performed through different optical devices: camera lenses, microscopes, remote fibre optic probe, etc. The size of the laser beam determines the surface analysed in one shot. In a macro configuration, 100 to 500 $\mu m^2$ are illuminated, typically. The laser spot is reduced to ~ 1 $\mu m^2$ for measurements with long focus, high magnification microscope objectives. Specific analysis of painting, glaze, underglaze décor, glaze/body interface can be made from the top using long focus microscope objectives (**Fig. 1**). On the other hand, body analysis can be performed on unglazed regions (rim, bottom or support spurs). A good technique is to map the sample surface using an XY motorised plater, hence selecting the most representative spectra of the different constituents [12]. The choice of the exciting radiation must be optimised. Blue or violet (short wavelength) excitation promotes a strong Raman scattering and provides a large spectral window with multichannel spectrographs, whereas red light must be preferred to analyse the low wavenumber region, below 200 $cm^{-1}$, specifically. The power of illumination needs to be reduced for coloured glazes if the excitation energy corresponds to the chromophore electronic absorption (resonance Raman spectra [11,14]), which induces local heating, and can modify the material through phase transition, oxidation, burning.

Sometimes, fluorescence is observed, for instance for excavated samples. The reliability of the technique has been extensively analyzed in [6].

## 2. Information extracted from the Raman spectra

The following information is present in Raman spectra:

i) phases nature : like for X-rays, identification of phases and polymorphs is the most easily obtained by comparison with data bases (see [10, 17] and refs herein). Note because the Raman cross section is directly related to the number of electron involved in the (covalent) bond, the technique efficiency is very variable: rare nanosized precipitates of coloured pigments are easily observed in Raman while they are not visible by XRD. On the other hand some main phases can be hardly observed (e.g. tridymite).

ii) the structure and composition of glassy phases. For instance Fig. 3 highlights the similarity between the glass of a Punic/Roman bead [4] and the glaze coating of a *Lâjvardina* ceramic [14]. Metal nanocrystals at the origin of lustre give low wavenumber modes (Fig. 1).

### 2.1 Identification from the spectra of crystalline phases

Typical examples of spectra obtained from the porcelain bodies are given in Fig. 2. The easy discrimination between soft- and hard-paste ceramics results from the high peak intensity of β-wollastonite ($CaSiO_3$, main band at ca. 970 $cm^{-1}$) and/or calcium phosphate (β-$Ca_3(PO_4)_2$, ca. 960 $cm^{-1}$) phases in soft-pastes, while these phases are absent in hard-paste [9,12]. On the other hand, mullite or mullite-like glassy phase spectra (main bands at ca. 480, 960 and 1130 $cm^{-1}$) are observed in high temperature fired hard-paste porcelains (prepared with kaolin and feldspars) [9-12]. Figure 2 shows representative Raman spectra recorded on the glazes from the three main factories in Paris area in the 18$^{th}$ century: Chantilly, Mennecy and Saint-Cloud. Identification is obvious from the comparison of the phases both in the body and in the glaze or at the glaze/body interface [12]. The Raman spectra obtained from modern and ancient Vietnamese celadon are compared Figure 3. The narrow peaks on modern celadon spectra correspond to crystalline $\alpha$-wollastonite, $CaSiO_3$, in the glaze (main bands at ca. 985 and 580 $cm^{-1}$). Examination of ancient celadons under a microscope shows that bubbles several tens of micrometers in size are almost homogeneously distributed in the glaze [11]. These micrometer size bubbles are an alternative to the precipitation of α-wollastonite to achieve a deep translucence. Chromium traces give green colour, whereas the jade-like colour is obtained from iron doping in the amorphous glaze of ancient celadons.

### 2.2 Identification from the spectra of amorphous silicates. Relationship between processing and Raman spectra

Glasses are silicate networks in which $SiO_4$ tetrahedra are bound together through oxygen atoms [21]. The properties are changed if these connections are modified by the incorporation of lead, alkali-earth /alkali elements, ... By comparing Raman spectra of glassy (or crystalline) silicates, it clearly appears that the Raman intensity of Si-O bending and stretching envelopes vary with composition [1-5,18,19]. Glass and glaze can be elaborated at temperatures ranging from ca 700°C (or less, for instance for pottery lustre) to 1450°C (high temperature fired porcelain glaze). A clear differentiation between the various glasses is possible, because the connectivity of the $SiO_4$ polymeric units can be investigated through the relative intensities of Si-O stretching and bending modes, at ca. 1000 and 500 $cm^{-1}$, respectively (**Fig. 3**) [2,19]. **Figure 4** compares the polymerisation index ($A_{500}/A_{1000}$ ratio with $A_{500}$, Raman area of the 500 $cm^{-1}$ band and $A_{1000}$, Raman area of the 1000 $cm^{-1}$ band) for a series of samples from Turkey (Byzantine to Seldjukide periods) [16]. The relationship between Raman index of polymerisation and both the glass composition and temperature of processing is well documented [2,13,19]: the first family ($A_{500}/A_{1000} < 0.3$) corresponds to most Islamic lead-containing or lustre potteries and rare Punic/Roman glasses; the second family (0.3 < $A_{500}/A_{1000}$ <0.8) consists of some 19th century lead-based soft-paste porcelain enamels and some Punic/Roman glasses (blue, green and colourless); the third family (0.8<$A_{500}/A_{1000}$ <1.1) corresponds to most ancient glasses and the 18th century soft-paste porcelain enamels; family #4 corresponds to celadon Ca-based enamels and family #5 to Ca-based porcelain enamels. Family #6 corresponds to K-based hard-paste porcelain glaze. The $A_{500}/A_{1000}$ ratio is strongly correlated to the processing temperature (~1400°C for $A_{500}/A_{1000}$ ~7, ~1000°C for $A_{500}/A_{1000}$ ~1 and ~ 600°C or less, for $A_{500}/A_{1000}$ ~0.3).
Because the $SiO_4$ tetrahedron is a very well defined vibrational and structural isolated entity, its different configurations have specific vibrational fingerprint. From the literature [1,6, 18] the different spectral components of the stretching envelope (Fig. 3) were assigned to the Si-O vibrations with zero ($Q^0$ or isolated $SiO_4$), one ($Q^1$ or dumbbell -$SiO_3$), two ($Q^2$ or =$SiO_2$), and three ($Q^3$ or ≡SiO) bridging oxygens per tetrahedral group, respectively. $Q^4$ corresponds to fully polymerised tetrahedra as in pure silica. Decomposition of the bending and stretching massifs is illustrated with the example of Vietnamese ceramics given in Fig. 3 [6]: a clear differentiation between two kiln productions from 13th and 15th century is obvious for $Q^1$, $Q^2$ and $Q^{3-4}$ components.

## 2.3 Pigments and resonance Raman spectroscopy

The size of the crystalline pigments dispersed in glassy coatings must be around to 100 nm to obtain a high-gloss glaze. Pigments are thus among the oldest materials for which the nanocrystalline state is tailored. For a given colouring ion, the result depends on the glaze composition, the firing temperature, the atmosphere and the grinding. In other words, the final colour depends on the technology used.
The different ways of colouring matrices will produce different Raman features [11]:
i) Dispersion of transition metal ions ($Cu^{2+}$, $Co^{2+ \text{ or } 3+}$, $Mn^{2+ \text{ or } 3+}$…) in the glass network of the coating. In this case, Raman scattering will not be very sensitive; ii) precipitation of "small" (coloured) crystals on cooling. This technique is widely used to opacify the matrix (e.g. cassiterite ($SnO_2$)). The small size of the crystals lead to a characteristic band-broadening in the Raman spectrum [11]; iii) dispersion of an insoluble coloured crystal (a pigment) in the coating matrix, with a characteristic Raman signature; iv) dispersion of metal nanoparticles like gold (Cassius purple) silver or copper (lustre ware)[10]. Very low wavenumber modes also appears and is characteristic of Cu/Ag metal precipitates at the origin of the metallic lustre. Note that because pigments are coloured, some resonance effects can occur, in which case Raman spectra will depend on the exciting light and the chromophore amount [10,11,14]. See, for instance, the spectrum of lapis lazuli in Fig. 4 : the fundamental bands related to $S_n$ chromophores at ca. 260 and 548 $cm^{-1}$ but also their harmonics ($n\nu$) and and combinations ($\delta+n\nu$) are visible.

## 3. Conclusion

Macro- and micro-Raman spectroscopy is a good tool for the non-destructive remote identification of glasses and ceramics. The Raman technique is efficient in spite of the small size, low crystallinity and low amount of the probed phases. A feasible discrimination has been demonstrated by considering the fingerprint of both crystalline and amorphous phases. Extracting some parameters from the Raman spectra of amorphous silicates allows their classification as a function of composition (Pb-, Ca- or K-based) and/or processing temperature. Spectra of amorphous matrices containing different coloured (nano)crystalline phases allows for the identification of the pigment as well as for recognition of artefacts prepared with the same/different technology.

# FIGURE CAPTIONS

**Fig. 1** : left, sketch of the Raman analysis of a ceramic : overglaze, underglaze, interphase, etc. Example of lutreware analyses down from the glaze surface : centre, Termez, Uzbekistan; right, excavated from Fostat, Egypt.

**Fig. 2** : Examples phases/technology relationships; *left*, in the body, mullite (M) for hard-paste porcelain (Meissen, Saxony, 19$^{th}$ century), β wollastonite (W) and calcium phosphate (P) for soft-paste porcelains (St-Cloud and Chantilly, France, 18$^{th}$ century); *centre*, in the glaze; *righ*t, comparison between a genuine Vietnamese celadon and a copy.

**Fig. 3** : *left*, comparison of spectra recorded on a Punic/Roman glass bead (ca. 2$^{nd}$ BC/AD, Carthage) and a blue glazed Lâjvardina ewer (13$^{th}$ century, Persia); *centre*, example of Si-O bending (ca. 500 cm$^{-1}$) and stretching (ca. 1000 cm$^{-1}$) massifs with $Q_n$ components for hard paste porcelain glazes with different silica content; *right*, centres of gravity and relative areas of the $Q_n$ components extracted from the Raman spectra recorded normal to the glaze surface (squares) and across the shard section (triangle) of the 13$^{th}$ century celadon (solids symbols) and the 15$^{th}$ century porcelain (open symbols).

**Fig. 4** *: right*, the polymerisation index is calculated from the area ratio of Si-O bending ($A_{500}$) and stretching ($A_{1000}$) envelopes ; the samples are glazes from Byzantine (6$^{th}$-15$^{th}$ centuries) and Seljukides (11-13$^{th}$ centuries) wares and tiles (from Efes, Iznik (I), Tokat (TT), Sivas (ST), Korucutepe, Turkey [16]); *right*, lapis lazuli resonance Raman spectrum recorded at the body/glaze interface in Lâjvardina ceramic (13$^{th}$, Persia, see Fig. 3).

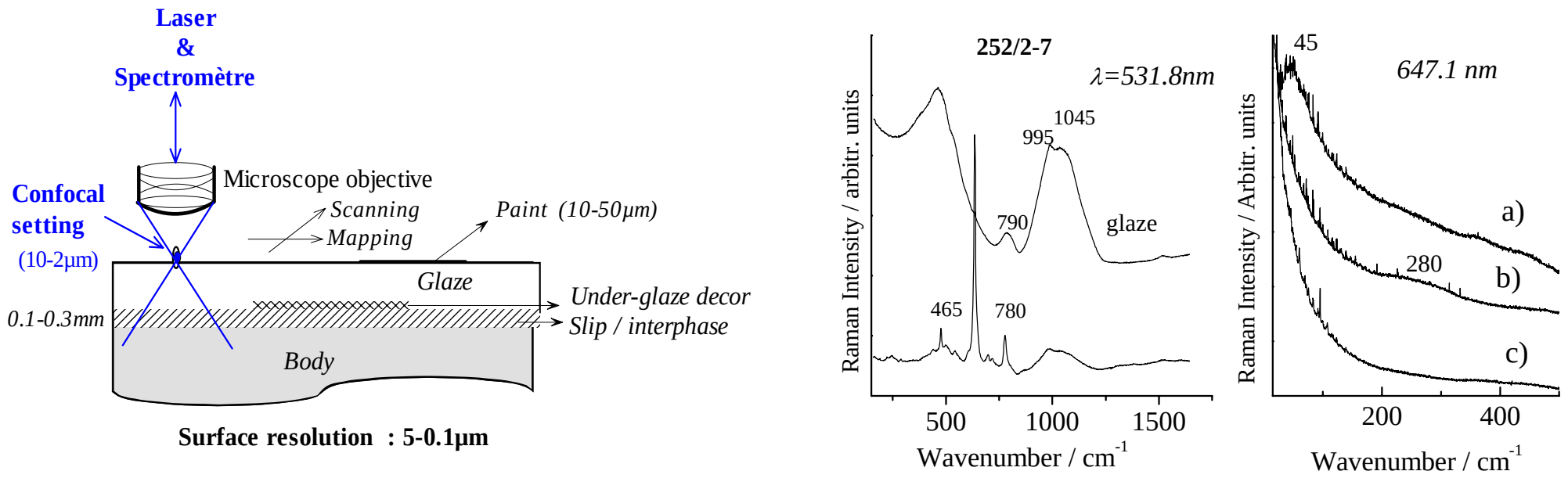

Fig. 1 :

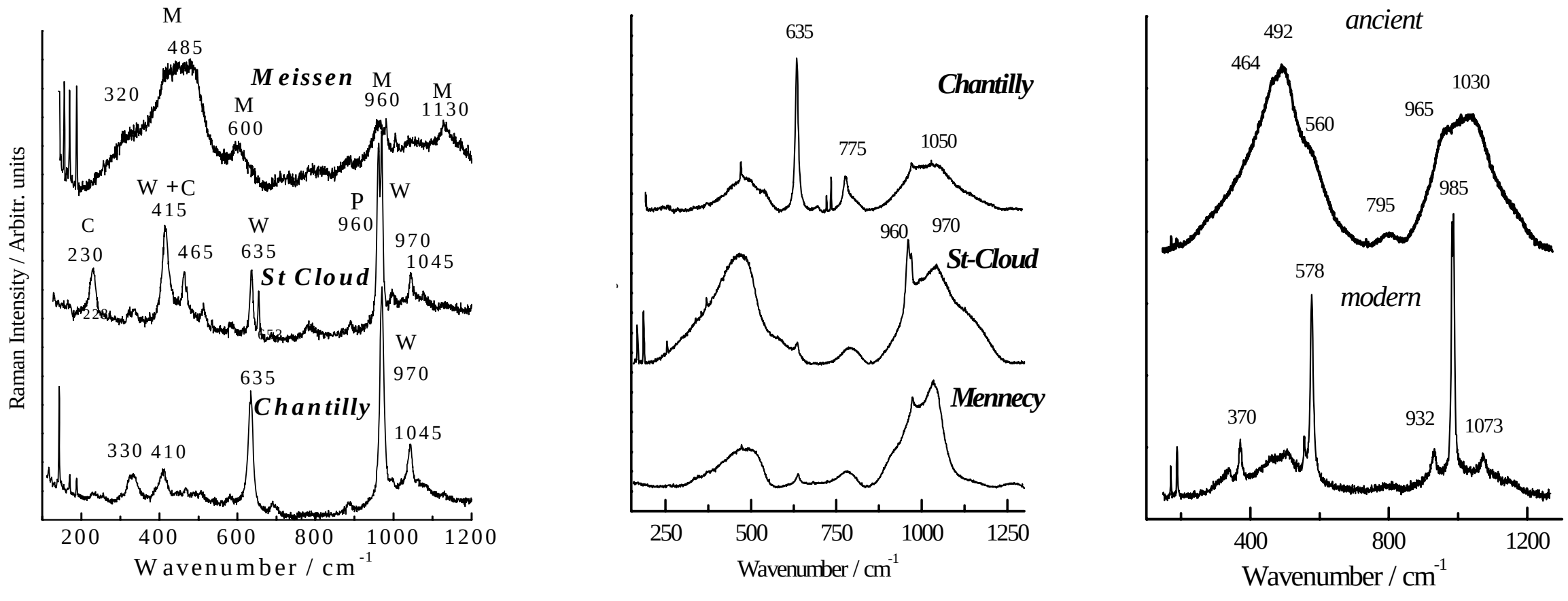

Fig. 2 :

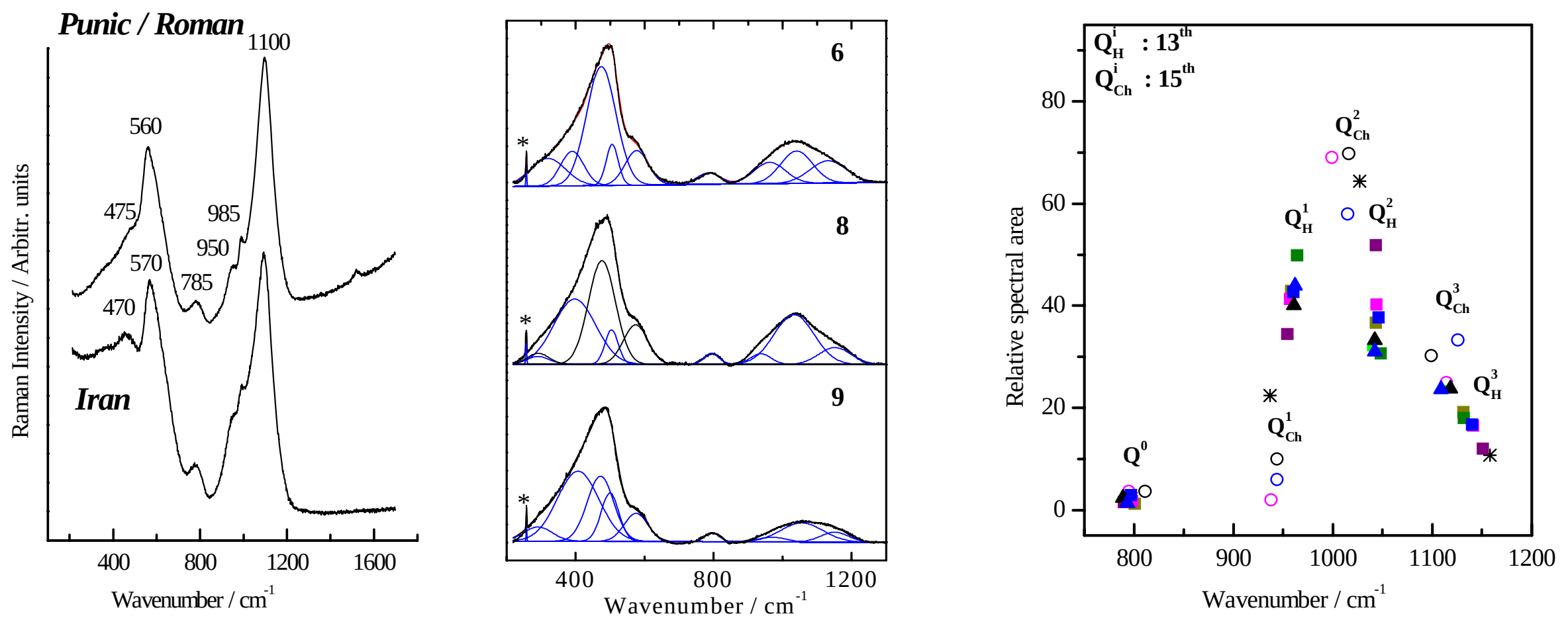

Fig. 3:

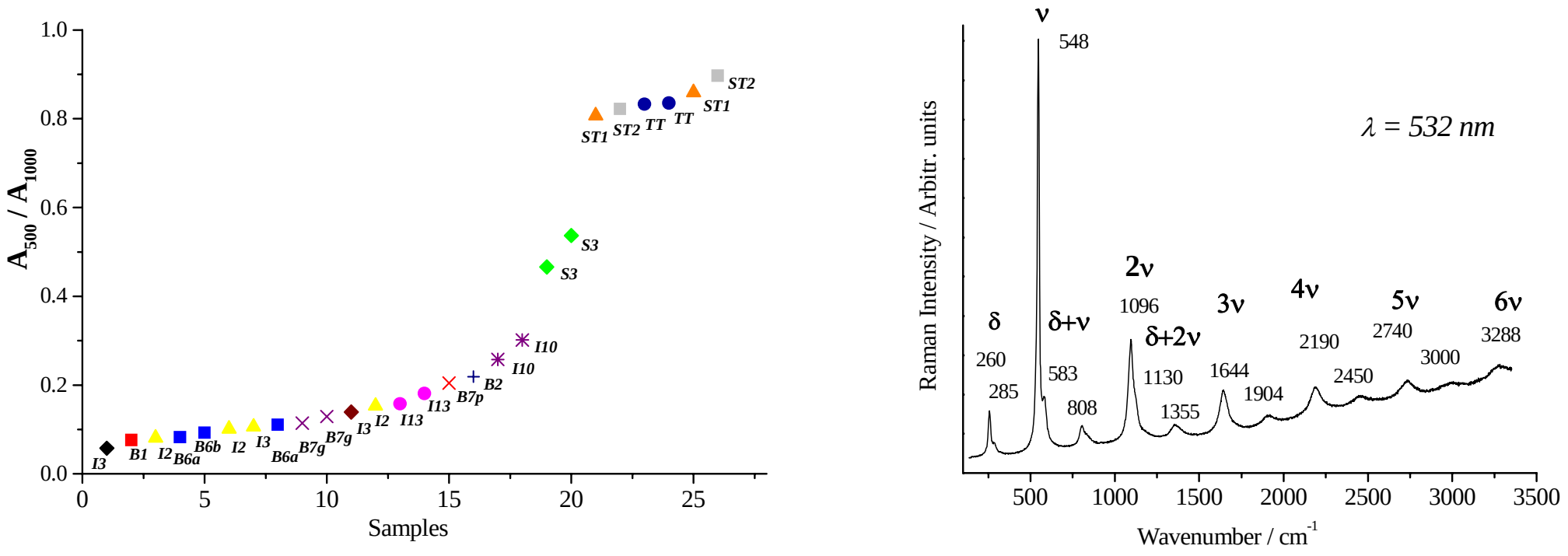

Fig. 4